\documentclass[CJK,a4paper,10pt,twoside]{cpc-hepnp}
\usepackage{upgreek,fancyhdr}
\usepackage{multicol}
\usepackage{graphicx}
\usepackage{booktabs}
\usepackage{amssymb,bm,mathrsfs,bbm,amscd}
\usepackage[tbtags]{amsmath}
\usepackage{lastpage}
\usepackage[bookmarks=true,colorlinks,linkcolor=blue,urlcolor=blue,citecolor=blue,plainpages=false,pdfpagelabels,final,breaklinks=true]{hyperref}
\usepackage{threeparttable,booktabs}
\usepackage{CJKutf8}
\begin{document}
\begin{CJK*}{GB}{gbsn}
%\begin{CJK*}{GBK}{song}

\title{A novel method of half-life determination for highly charged ions based on the isochronous mass spectrometry}
\author{%	
Qi Zeng$^{1,2}$
\quad Bushi Huang$^{1,2}$
\quad Tingwei Peng$^{1,2}$
\quad Hongfu Li $^{3}$\email{lihongfu@impcas.ac.cn.}%
\quad Xing Xu$^{3,4}$\email{xuxing@impcas.ac.cn.}\\%
\quad Yuanming Xing$^{3,4}$
\quad Huaiqiang Zhang$^{1,2}$
\quad Jiankun Zhao$^{1,2}$
\quad Xu Zhou$^{3,4}$
}
\maketitle

\address{%
$^1$ Jiangxi Province Key Laboratory of Nuclear Physics and Technology, East China University of Technology, Nanchang 330013, China\\
$^2$ Jiangxi Engineering Technology Research Center of Nuclear Radiation Detection and Application, Nanchang 330013, China\\
$^3$ State Key Laboratory of Heavy Ion Science and Technology, Institute of Modern Physics, Chinese Academy of Sciences, Lanzhou 730000, China\\
$^4$ School of Nuclear Science and Technology, University of Chinese Academy of Sciences, Beijing 100049, China
}

\begin{abstract}
The lifetime of the isomeric state in fully stripped $^{94}$Ru$^{44+}$ ions has been measured using isochronous mass spectrometry (IMS) at the experimental Cooler Storage Ring (CSRe) of the Heavy Ion Research Facility in Lanzhou (HIRFL).
Previously, the isomeric lifetime was determined by analyzing the decay time points of individual decay events.
In this paper, we present a novel approach to determine the isomeric lifetime based on the survival time of ions obtained from IMS.
The survival lifetime of the ground and isomeric states of $^{94}$Ru$^{44+}$ were measured to be $270(9)\,\mu$s and $121(4)\,\mu$s in the laboratory, respectively.
Given that the ground state of $^{94}$Ru$^{44+}$ has a natural lifetime of approximately 75 min, its survival lifetime in the experimental setup was predominantly determined by the beam-loss lifetime, including interactions with residual gas in the storage ring and carbon foil of the detector. In contrast, the survival lifetime of $^{94m}$Ru$^{44+}$ was governed by its intrinsic nuclear lifetime and additional beam-loss effects.
The nuclear decay lifetime of $^{94m}$Ru$^{44+}$ was extracted through differential survival lifetime analysis between ground and isomeric state, under the assumption that the beam-loss lifetimes for both quantum systems are identical.
Using this novel methodology, the lifetime measured in laboratory was $221(14)\,\mu$s. After relativistic time-dilation corrections, the corresponding rest-frame half-life was calculated to be $118(7)\,\mu$s. This result demonstrates excellent consistency with previous experimental results, validating the reliability of the new method.
This method is suitable for determining half-lives of highly charged ions in the range of several tens of microseconds to milliseconds using IMS.
\end{abstract}


\begin{multicols}{2}
	
\section{Introduction}

The decay characteristics of highly charged ions (HCIs) serve as sensitive probes for nuclear-electron coupling effects, tests of collective nuclear models, and constraints for astrophysical nucleosynthesis pathways in stellar environments~\cite{Schatz,Cowan,Langanke}.
In bare ions, the absence of atomic electrons eliminates competing decay channels such as internal conversion (IC) and orbital electron capture (EC), permitting direct measurement of $\gamma$-decay branching ratios~\cite{Attallah,Harston,Litvinov2003,Reed,Zeng2017}.
Therefore, HCIs offer a unique approach for investigating the nuclear structure of excited states.

Storage-ring based experiments employing Schottky mass spectrometry (SMS) has been successfully leveraged for investigating long-lived isomers ($\tau > 1$ s) through characteristic frequency shifts~\cite{Litvinov2004,Reed2010,Reed2012,Sidhu,Leckenby}, particularly confirming the existence of high-K isomers. These studies provided direct evidence for nuclear deformation effects in heavy nuclei.
A milestone achievement was the lifetime measurement for hydrogen-like $^{192m}$Os$^{75+}$, where the observed lifetime extension in relation to that of neutral atoms validated relativistic Dirac-Fock calculations of internal conversion coefficients in HCIs~\cite{Akber,Band}.
Recent developments in Schottky-isochronous mass spectrometry (S+IMS)~\cite{Tu2018,Freire} have reduced the measurable half-life threshold down to approximately 24 ms~\cite{Freire}. However, microsecond-scale decays remain challenging owing to necessary electron cooling  and limited signal-to-noise ratio of the Schottky resonator.

To overcome these challenges, an innovative method for identifying the in-ring decay using isochronous mass spectrometry (IMS) was proposed in the experimental storage ring of the Heavy Ion Research Facility in Lanzhou (HIRFL-CSR)~\cite{Zeng2017}. In that experiment, a sudden change in the revolution time of fully stripped $^{94m}$Ru$^{44+}$ was recognized as the fingerprint of the isomer decay when stored in the ring. The advantage of this method is that the revolution times of stored ions in an IMS experiment are measured in time intervals shorter than $\mu$s. Consequently, it is sensitive enough to observe nuclear decays occurring on time scales ranging from a few $\mu$s to a few hundred $\mu$s, establishing $^{94m}$Ru$^{44+}$ as the shortest-lived nuclear state with directly measured mass.

However, in previous data analyses, only decay events of $^{94m}$Ru$^{44+}$ were used, leading to a relatively large error of the determined half-life. In this study, we propose a refined methodology to determine the half-life of $^{94m}$Ru$^{44+}$ using beam-loss constants individually evaluated for its ground and isomeric states. In this new method, all events of the decay and non-decay isomers are utilized, thereby enhancing the precision of half-life
determinations for short-lived highly charged ions (HCIs)
using IMS.

\section{Experimental Setup}

The experiment was performed at the Heavy Ion Research Facility in Lanzhou (HIRFL)~\cite{Xia}. A primary beam of $^{112}$Sn$^{35+}$ was accelerated to an energy of 376.42~MeV/u with an intensity of 7$\times10^7$ particle per pulse, and subsequently fast-extracted to impinge on a $^9$Be target located at the entrance of the fragment separator of the Radioactive Ion Beam Line in Lanzhou (RIBLL2). The resulting projectile fragments were selected and purified via RIBLL2, and a carbon stripper foil placed at the exit further ionized the fragments before they were injected into the experimental Cooler Storage Ring (CSRe).

For the optimization of IMS conditions, the CSRe was adjusted to a transition point of $\gamma_t = 1.302$ with a magnetic rigidity of $B\rho = 5.5294$~Tm. The storage ring was operated in the isochronous mode, ensuring a revolution time nearly independent of the velocity of the ions for the nuclei of interest, $^{94}$Ru$^{44+}$.

The revolution times of the stored ions were measured using a time-of-flight (TOF) detector based on a micro-channel plate (MCP)~\cite{Mei2010}. The TOF detector incorporated a carbon foil with a thickness of approximately 19~$\mu$g/cm$^2$ and a diameter of 40~mm, mounted at the geometric center of the beam line. As ions passed through the carbon foil, secondary electrons were emitted. These electrons were accelerated by an electric field (130~V/mm) and deflected by a perpendicular magnetic field ($\sim$80~Gs) toward the MCP detector. Upon reaching the MCP, the electrons were amplified, producing signals that were transmitted through high-frequency coaxial cables to a Tektronix DPO71254 digital oscilloscope operating at a
sampling rate of 50 GHz for offline analysis. Additional experimental details are available in Ref.~\cite{Zeng2017}.

\section{Data analysis}
For each particle circulating in the ring, a time sequence
composed of the time stamps when passing the TOF detector as a function of the revolution number was extracted from the recorded signals.
As mentioned in a previous study\cite{Zeng2024}, only the isomeric states decayed within the observation window [15 $\mu$s, 185 $\mu$s ] can be identified. To determine the beam lost constant,  all ions that circulated for more than 15 $\mu$s were considered in the analysis based on the procedures described in Ref. \cite{Mei2010,tu2011,chen2017}.
In the revolution spectrum, the events with revolution times between 670.90 and 670.98 ns were identified as $^{94}$Ru$^{44+}$.
All these events can be classified into three categories: ground, isomeric, and decayed isomeric states.
The decayed isomeric state event refers to cases where the isomer decays to the ground state within the observation window.
After decay, the ion continues circulating in the CSRe as a ground state.
These decayed events were identified using the method described in Ref.~\cite{Zeng2024}.
Next, we need to identify the remaining events as either ground state or isomeric state.
According to the excitation energy of $^{94m}$Ru and the optical parameters of the CSRe, the revolution time difference between the isomeric and ground states was 11.9~ps.

\begin{center}
\includegraphics[width=9cm]{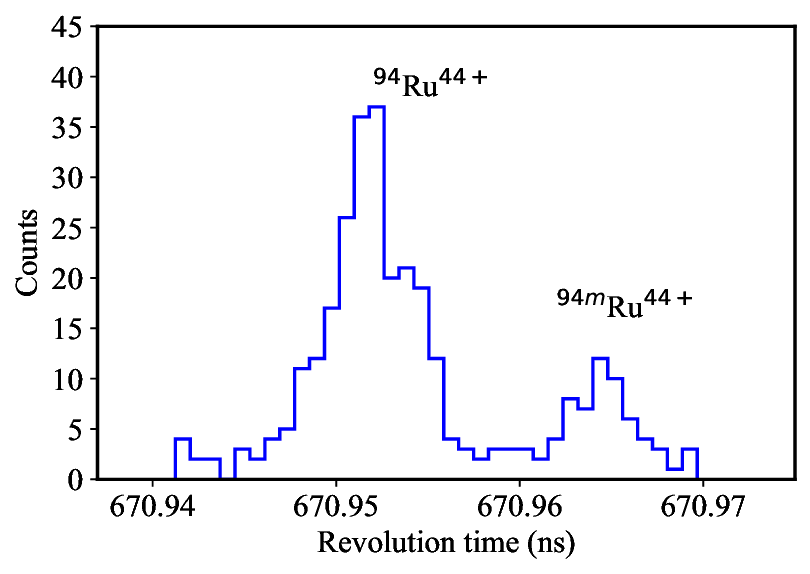}
\figcaption{\label{fig1}Revolution time spectrum of the ground and isomeric states of $^{94}$Ru$^{44+}$  after magnetic field correction.}
\end{center}

In the revolution time spectrum obtained by directly accumulating revolution times into a histogram, the ground and isomeric states of $^{94}$Ru$^{44+}$ cannot be resolved owing to magnetic field instabilities, which also induce shifts in revolution times across different injections.
By leveraging the fact that multiple ions are stored simultaneously in the CSRe and assuming identical shifts for all ions within a single injection, we applied the method described in Ref.~\cite{Xing} to correct for magnetic field instabilities.
This correction resulted in a higher-resolution revolution period spectrum, as shown in Fig.~\ref{fig1}.
The events whose revolution times were smaller than 670.958~ns are marked as ground states, whereas the remaining events are identified as isomeric states.

\begin{center}
\includegraphics[width=8cm]{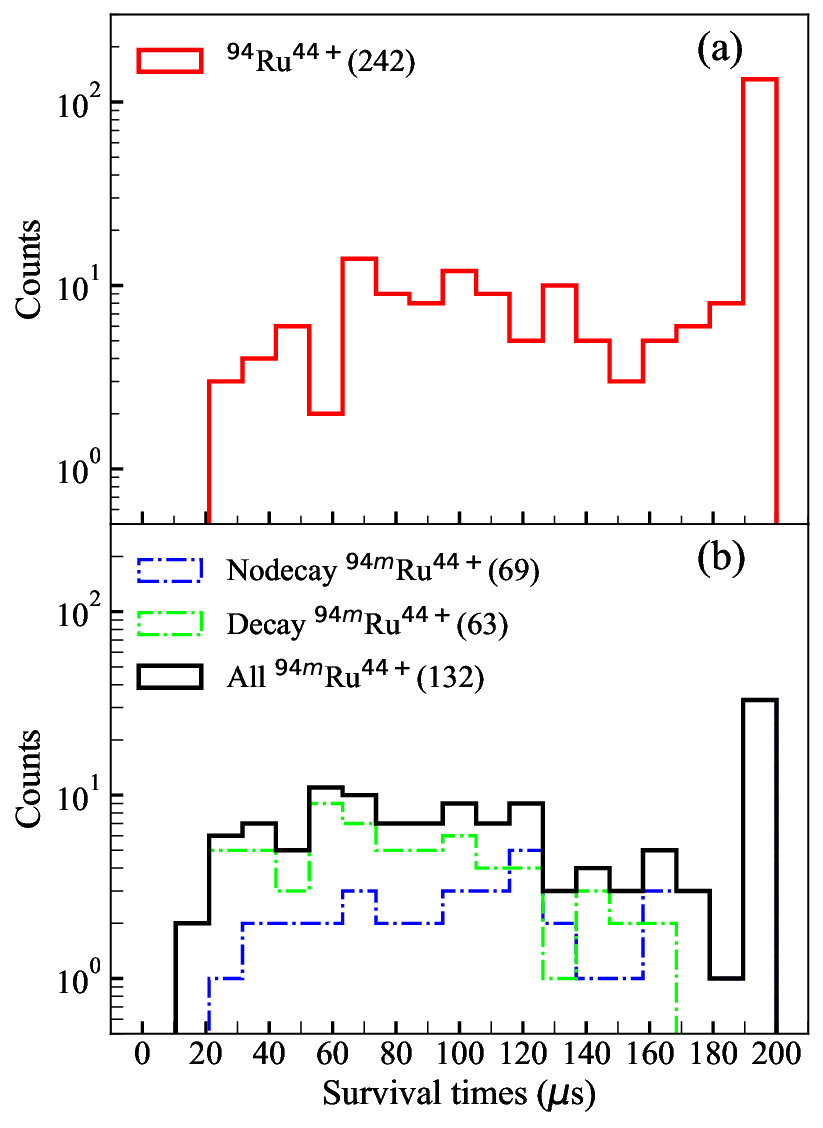}
\figcaption{\label{fig2}Distribution of the last timing signal of the ions for the (a) ground and (b) isomeric states of $^{94}$Ru$^{44+}$. The numbers in parentheses within the legend denote the counts of corresponding events.  }
\end{center}

For each ion circulating in the IMS, the TOF detector tracked it turn by turn.
The last timing signal of each ion in the observation window was marked as the decay point for the ground state of $^{94}$Ru.
Given that the half-life of the ground state of $^{94}$Ru is approximately 75 min, which is much longer than the observation window, the corresponding events would be lost due to non-radiation decay, the interaction with the residual gas in the CSRe and the carbon foil of the TOF detector.
The last timing signal was set as the decay point for these events.
The distribution of the survival time of the ground state of $^{94}$Ru is shown in Fig. \ref{fig2}(a).
For events in the isomeric state, decay could occur through both the radiation, gamma decay, and non-radiation processes.
For events that could still circulate in the CSRe after decay, the decay time point was determined using the approach described in ref\cite{Zeng2024}.
For the remaining isomeric events, the decay time point in the observation windows was also set as the last timing signal.
The distribution of the survival time of the isomeric state of $^{94}$Ru is shown in Fig. \ref{fig2}(b).

\begin{center}
\includegraphics[width=9.5cm]{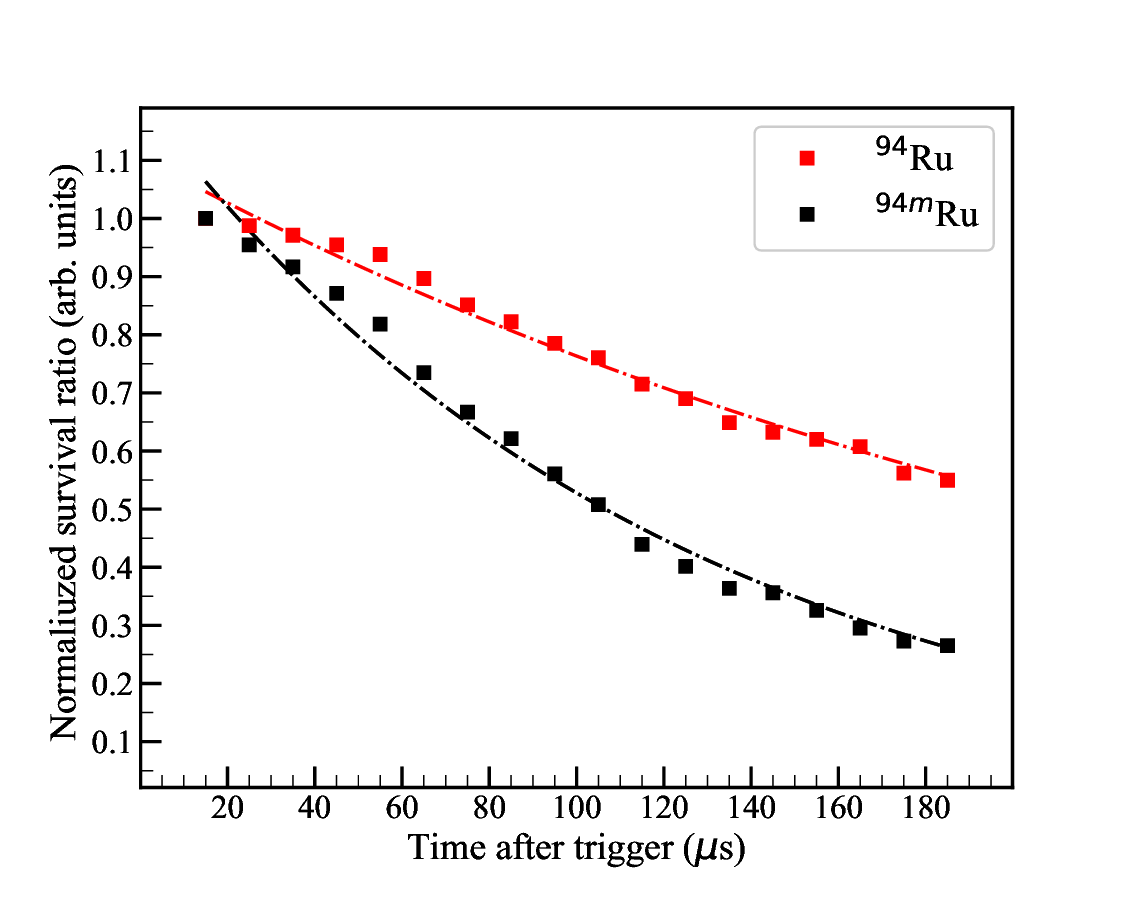}
\figcaption{\label{fig3} Normalized survival ratios of the  ground and isomeric states of $^{94}$Ru$^{44+}$ after the trigger of DAQ. The dash-dotted lines represent the results of exponential function fitting. }
\end{center}

The normalised survival ratio (R) for the i$^{th}$ bin ground (isomeric) state of $^{94}$Ru is defined as,
\begin{eqnarray}
\label{eq0}
R_i = \frac{Sum-\sum_{j = 1}^{i-1} Count_{j}}{Sum},
\end{eqnarray}
where $Sum$ is the total count of the  ground (isomeric) state and $Count_{j}$ is the count in the j$^{th}$ bin in Fig.\ref{fig2}.
Given that the observation window starts at 15$\mu$s after the trigger of DAQ, the normalised survival ratio at 15 $\mu$s was set as 1.
The normalised survival ratio as a function of the time after the trigger is shown in Fig\ref{fig3}.
The decay constants for both the ground and isomeric states of $^{94}$Ru, $\lambda_{g.s.}$ and $\lambda_{i.s.}$, were separately determined by fitting the normalised survival ratio as a function of the time after the trigger, T, with a single exponential function,
\begin{eqnarray}
\label{eq1}
R = R_0e^{-\lambda _{g.s(i.s)T}},
\end{eqnarray}
and the corresponding values of goodness of fit are 0.983 for the ground state and 0.984 for the isomeric state.
The decay constant of the ground state, $\lambda_{\text{g.s.}}$, was determined to be 0.00371(12)~$\mu$s$^{-1}$, while that of the isomeric state, $\lambda_{\text{i.s.}}$, was measured to be 0.00824(26)~$\mu$s$^{-1}$. Accordingly, the survival lifetimes of the ground and isomeric states of $^{94}$Ru$^{44+}$ in the laboratory were deduced to be $270(9)\,\mu$s and $121(4)\,\mu$s, respectively.
The decay constant of $^{94m}$Ru, $\lambda_{i.s.}$  is a sum of three components: internal conversion (IC) decay constant $\lambda_{IC}$, $\gamma$ decay constant $\lambda_{\gamma}$, and beam-loss constant $\lambda^*_{loss}$ due to collisions with residual gas atoms or the carbon foil in the TOF detector,
\begin{eqnarray}
\label{eq1}
\lambda_{i.s.} = \lambda_{IC}+\lambda_{\gamma}+\lambda_{loss}
\end{eqnarray}
It is evident that for the fully stripped ion $^{94m}$Ru, $\lambda_{IC}$ = 0 owing to the absence of bound electrons.
The beam loss constants depend typically only on Z, therefore, the beam-loss constant for ground and isomeric states of $^{94}$Ru would be identical.
As previously mentioned, the decay half-life of the ground state of $^{94}$Ru is 75 min, which is much longer than its survival time in the CSRe.
Thus, $\lambda_{loss}$ would be equal to $\lambda_{g.s}$.
Then, $\lambda_{\gamma}$ for $^{94m}$Ru was calculated to be 0.00453(28) $\mu$s$^{-1}$.
Taking into account the Lorentz factor, $\gamma$ = 1.302, deduced from the magnetic rigidity of the CSRe, T$_{1/2}$($^{94m}$Ru$^{44+}$) = 1/$\lambda_{\gamma}$ $\times$ ln(2) /$\gamma$ =  118(7) $\mu s$ in the rest frame, which is in good agreement with previous measurements.

\section{Summary}

The survival time of ions stored in the experimental Cooler Storage Ring (CSRe) can be determined through isochronous mass spectrometry (IMS). By analyzing the distribution of survival ratios within a defined observation window, the decay constant can be extracted with relatively high precision.

In this study, we determined the beam loss constants for both the ground and isomeric states of $^{94}$Ru. The beam loss associated with the ground state was dominated by nonradioactive processes. 
The non radioactive loss rate for the isomeric state was assumed to be the same as that for the ground state.
By subtracting the nonradioactive component from the total beam loss constant of the isomer, we derived the half-life of fully stripped $^{94m}$Ru$^{44+}$ as $118(7)\,\mu$s.
This result is in good agreement with the previously reported value of $102(17)\,\mu$s~\cite{Zeng2017}, with half the uncertainty of the earlier value. This substantial improvement confirms both the validity and enhanced reliability of the revised analysis method presented in this paper.

\end{multicols}

\begin{multicols}{2}

\end{multicols}

\clearpage
\end{CJK*}
\end{document}